\documentclass[useAMS,usenatbib]{mn2e}

\usepackage[dvips]{graphicx}

\def \aap {A\&A }
\def \apj {ApJ }

\def \mnras {MNRAS }

\def \etal {\textit{et al. }}
\def \corot {\textit{CoRoT }}
\def \kepler {\textit{Kepler }}

\title[The SARS algorithm]{The SARS algorithm: detrending \textit{CoRoT} light curves with Sysrem using simultaneous external parameters}

\author[A. Ofir]{Aviv Ofir$^{1}$\thanks{E-mail: avivofir@wise.tau.ac.il}, Roi Alonso$^{5,19}$, Aldo Stefano Bonomo$^{5}$, Ludmila Carone$^{18}$, Stefania Carpano$^{2}$,
\newauthor Benjamin Samuel$^{6}$, J\"org Weingrill$^{17}$, Suzanne Aigrain$^{3,4}$, Michel Auvergne$^{23}$,
\newauthor Annie Baglin$^{23}$, Pierre Barge$^{5}$, Pascal Borde$^{6}$, Francois Bouchy$^{7,8}$, Hans J. Deeg $^{9,10}$,
\newauthor Magali Deleuil$^{5}$, Rudolf Dvorak$^{12}$, Anders Erikson$^{13}$, Sylvio Ferraz Mello$^{11}$,
\newauthor Malcolm Fridlund$^{14}$, Michel Gillon$^{19,22}$, Tristan Guillot$^{15}$, Artie Hatzes$^{16}$, Laurent Jorda$^{5}$, 
\newauthor Helmut Lammer$^{17}$, Alain Leger$^{6}$, Antoine Llebaria$^{5}$, Claire Moutou$^{5}$, Marc Ollivier$^{6}$, 
\newauthor Martin P\"aetzold$^{18}$, Didier Queloz$^{19}$, Heike Rauer$^{13,20}$, Daniel Rouan$^{5}$, Jean Schneider$^{21}$, 
\newauthor and  Guenther Wuchterl$^{16}$\\
$^{1}$School of Physics and Astronomy, Raymond and Beverly Sackler Faculty of Exact Sciences, Tel Aviv University, Tel Aviv, Israel\\
$^{2}$RSSD, ESTEC/ESA, PO Box 299, 2200 AG Noordwijk, The Netherlands.
$^{3}$ School of Physics, University of Exeter, Exeter, EX4 4QL\\
$^{4}$ Oxford Astrophysics, University of Oxford, Keble Road, Oxford OX1 3RH, UK\\
$^{5}$ LAM, UMR 6110, CNRS/Univ. de Provence, 38 rue F. Joliot- Curie, 13388 Marseille, France\\
$^{6}$ Institut d'Astrophysique Spatiale, Universite Paris XI, F-91405 Orsay, France\\
$^{7}$ Institut d'Astrophysique de Paris, UMR7095 CNRS, Universit´e Pierre \& Marie Curie, 98bis boulevard Arago, 75014 Paris, France\\
$^{8}$ 22 Observatoire de Haute-Provence, CNRS/OAMP, 04870 St Michel l'Observatoire, France\\
$^{9}$ Instituto de Astrof´ýsica de Canarias , E-38205 La Laguna, Tenerife, Spain\\
$^{10}$ Dept. de Astrofsica, Universidad de La Laguna, Tenerife, Spain\\
$^{11}$ IAG Universidade de Sao Paulo, Sao Paulo, Brasil\\
$^{12}$ Institute for Astronomy, University of Vienna, T\"urkenschanzstrasse 17, 1180, Vienna, Austria\\
$^{13}$ Institute of Planetary Research, DLR, 12489 Berlin, Germany\\
$^{14}$ Research and Scientific Support Department, European Space Agency, Keplerlaan 1, NL-2200AG, Noordwijk, The Netherlands\\
$^{15}$ Universit de Nice Sophia Antipolis, CNRS, Observatoire de la Cte d'Azur, BP 4229, 06304 Nice, France\\
$^{16}$ Th\"uringer Landessternwarte, 07778 Tautenburg, Germany\\
$^{17}$ Space Research Institute, Austrian Academy of Science, Schmiedlstr. 6, A-8042 Graz, Austria\\
$^{18}$ Rheinisches Institut f\"ur Umweltforschung an der Universit\"at zu K\"oln, Aachener Strasse 209, 50931, K\"oln, Germany\\
$^{19}$ Observatoire de Gen`eve, Universit´e de Gen`eve, 51 chemin des Maillettes, 1290 Sauverny, Switzerland\\
$^{20}$ TU Berlin, Zentrum f\"ur Astronomie und Astrophysik, Hardenbergstr. 36, 10623 Berlin, Germany\\
$^{21}$ LUTH, Observatoire de Paris-Meudon, 5 place Jules Janssen, 92195 Meudon, France\\
$^{22}$ IAG Universit´e du Li`ege, All´ee du 6 auˆot 17, Li`ege 1, Belgium\\
$^{23}$ LESIA, Observatoire de Paris-Meudon, 5 place Jules Janssen, 92195 Meudon, France
}

\begin{document}

\date{Submitted...}

\pagerange{\pageref{firstpage}--\pageref{lastpage}} \pubyear{2002}

\maketitle

\label{firstpage}

\begin{abstract}
Surveys for exoplanetary transits are usually limited not by photon noise but rather by the amount of red noise in their data. In particular, although the \corot space-based survey data are being carefully scrutinized, significant new sources of systematic noises are still being discovered. Recently, a magnitude-dependant systematic effect was discovered in the \corot data by Mazeh \& Guterman \etal and a phenomenological correction was proposed. Here we tie the observed effect a particular type of effect, and in the process generalize the popular Sysrem algorithm to include external parameters in a simultaneous solution with the unknown effects. We show that a post-processing scheme based on this algorithm performs well and indeed allows for the detection of new transit-like signals that were not previously detected.
\end{abstract}


\section{Introduction} \label{Intro}
The limiting factor for most planetary transit surveys is not the theoretical photon noise but rather the practically-achieved red noise from non-astrophysical sources (Pont \etal 2006). The most capable transit surveys are the space-based surveys \corot \footnote{The CoRoT space mission, launched on December 27th 2006, has
been developed and is operated by CNES, with the contribution of Austria, Belgium, Brazil, ESA, Germany, and Spain. \corot data become publicly available one year after release to the Co-Is of the mission from the \corot archive: \tt{ http://idoc-corot.ias.u-psud.fr/}.} and \kepler because their stable environments allow for minimal red noise (and other benefits - such as continuous observations). Still, no instrument is perfect and the \corot light curves are known to show a number of significant effects that hinder transit detection, among them: discontinuities of arbitrary magnitude due to high energetic protons flux near the South Atlantic Anomaly (SAA), residuals at the \corot orbital period, spacecraft jitter, CCD long-term aging and more. Many of these effects are correctable to satisfactory level at post-processing, but not for all stars and at all times.

On top of these effects, Mazeh \& Guterman \etal (2009) (hereafter MG09) recently discovered that there are significant magnitude-dependant systematic effects in the \corot light curves, and they developed a phenomenological algorithm to correct for them. In this paper we tie the above effect to a particular type of effect: added/subtracted linear flux, and are thus able to improve on their correction. In the process we generalize the Sysrem algorithm (Tamuz \etal 2005) to include arbitrary external parameters and show the benefits of using this modified version. Below, we formulate the Sysrem generalization in \S \ref{SARScore}, which is part of a complete post-processing scheme presented in \S \ref{SARSpipeline}, and conclude.

\section{The SARS core}
\label{SARScore}

\subsection{Algorithm}

MG09 first noted that there are magnitude-dependent systematic effects in the \corot data. They proposed to correct for the effects by fitting a parabola to the residuals of each exposure --- but this correction is a purely phenomenological correction since there is no identified cause for the effects, and thus no explanation
as to why a parabola is the best functional form. We hypothesize that the underlying physical mechanism MG09 were trying to correct for is a constant flux that is either added or subtracted from all the light curves due to calibration errors, scattered light, or other causes. Such an additive effect will create a large magnitude difference on faint stars, and small magnitude difference on bright stars, as MG09 had originally observed. Indeed, the original authors had also considered this option (Tsevi Mazeh, personal comm.) but they chose to use a more phenomenological correction rather then to tie the correction to this proposed physical mechanism. Since detrending algorithms can't a priori disentangle additive from relative effects, we choose to simultaneously correct both types of effects, and so developed ``Simultaneous Additive and Relative Sysrem" -- or the SARS algorithm -- described below.

Suppose a matrix of photometric measurements of $N$ stars ($i=1...N$) on $M$ measurements ($j=1...M$) is given, so that the magnitude value of the $i^{th}$ star on the $j^{th}$ frame is $m_{ij}$ and its associated error is $\sigma_{i,j}$. After removing from each stellar light curve (hereafter LC) its mean or median $\bar{m}_i$ we are left with the matrix of residuals $r_{ij}$. In the original Sysrem, the residuals of intrinsically-constant stars are modeled using two contributions:

\begin{equation}
r_{ij}=A_j C_i + noise
\end{equation}

Where $A_j$ is the effect in each exposure and $C_i$ is effect's coefficient for each star. We note that since the data is in the magnitude system the effects found in this manner are relative in flux. In order to test out hypothesis that the magnitude-dependent effects stem from something that is additive in flux, we introduce the SARS model:

\begin{equation}
r_{ij}=A_j x_{ij} C_{A,i} + R_j C_{R,i} + noise
\label{SARSmodel}
\end{equation}

Here the second term is exactly the usual Sysrem effect were we simply change the letter from $A_j$ to $R_j$ to designate that it is a relative effect. The first term stands for the additive effect by introducing $x_{ij}=10^{0.4m_{ij}}$ which makes sure that the effect is stronger for faint stars and weaker for bright stars -- and in the correct functional form expected from additive flux effects. In practice, we use    $x_{ij}=10^{0.4(m_{ij}-m_{rel})}$ where $m_{rel}$ is a constant number (e.g., the median of all the stars on all the exposures) to avoid overly -large or -small values for $x_{ij}$. As in Sysrem, minimizing the sum of squared residuals $S$

\begin{equation}
S=\sum \bigg(\frac{r_{ij}-model}{\sigma_{ij}}\bigg)^2
\end{equation}

Gives the the besf-fitting effects $R$ and $A$, and the corresponding coefficients $C_R$ and $C_A$, which are:

\begin{equation}
A_j=\frac{\sum_{i} \frac{{r_{ij} x_{ij} C_{A,i}}}{\sigma_{ij}^2}  -   R_j \sum_{i} \frac{{C_{R,i} x_{ij} C_{A,i}}}{\sigma_{ij}^2}}{\sum_{i} \frac{{x_{ij}^2 C_{A,i}^2}}{\sigma_{ij}^2}}
\label{Aj}
\end{equation}

\begin{equation}
R_j=\frac{\sum_{i} \frac{{r_{ij} C_{R,i}}}{\sigma_{ij}^2}  -   A_j \sum_{i} \frac{{C_{R,i} x_{ij} C_{A,i}}}{\sigma_{ij}^2}}{\sum_{i} \frac{{C_{R,i}^2}}{\sigma_{ij}^2}}
\label{Rj}
\end{equation}

\begin{equation}
C_{A,i}=\frac{\sum_{j} \frac{{r_{ij} x_{ij} A_j}}{\sigma_{ij}^2}  -   C_{R,i} \sum_{j} \frac{{A_j x_{ij} R_j}}{\sigma_{ij}^2}}{\sum_{j} \frac{{x_{ij}^2 A_j^2}}{\sigma_{ij}^2}}
\label{CAi}
\end{equation}

\begin{equation}
C_{R,i}=\frac{\sum_{j} \frac{{r_{ij} R_j}}{\sigma_{ij}^2}  -   C_{A,i} \sum_{j} \frac{{A_j x_{ij} R_j}}{\sigma_{ij}^2}}{\sum_{j} \frac{{R_j^2}}{\sigma_{ij}^2}}
\label{CRi}
\end{equation}

As in Sysrem, the values of $A_j$, $R_j$, $C_{A,i}$, $C_{R,i}$ are iteratively refined until convergence is achieved. We note that we found that it is important that in each iteration the new values of the effects are used to calculate the coefficients, and not the values of the previous iteration. 

\textbf{Further generalization:} The formulae \ref{Aj} to \ref{CRi} do not ``know" that $x_{ij}$ is meant to scale magnitude data to create flux-based correction -- they only know that $x_{ij}$ depends on external information not present in the orignial matrix of residuals. In this, SARS present a significant departure from the original Sysrem by allowing for the detrending against any explicitly known external parameters as long as their effect can be encapsulated in some $x_{ij}$. For example, these can be: distance from the center of the CCD or pixel phase (or otherwise location-based), CCD temperature (or otherwise weather related), or Moon phase (or otherwise temporal effects), etc. It is thus easy to include multiple external effects in the detrending model (e.g. the SARS model of eq. \ref{SARSmodel}), and by minimizing the sum of squared residuals $S$ to establish their effect on the data simultaneously with the effects of unknown sources.

\subsection{Suggested Good Practices}

Below we describe what we think are good practices when using SARS:
\begin{itemize}
\item \textbf{Starting point:} We note that already the Sysrem ``search space" was very large: as many parameters to adjust as there are stars + exposures. By simultaneously fitting more than one effect in SARS -- we further enlarge this search space greatly. In the original Sysrem the starting point was deemed to be unimportant since Tamuz \etal (2005) claimed that in their simulations no matter what initial values were used, the same effect and coefficients were obtained. We believe that these simulations were somewhat lacking in that they used white noise only, with no red noise component, which allowed them to always find the (unique) global $\chi^2$ minimum with no local minima to be avoided. We have also performed a similar test -- but on real data, rather than simulated data, and found that sometimes (a few percent of the runs) the global minimum was indeed missed. Fearing that the enlarged SARS search space would worsen the problem, we choose to start the iterations from a deterministic point. Assuming that the median of photometric measurements is rather robust, we start by finding a proxy to the relative and additive effects by: $R_j$=median($r_{ij}$) and $A_j$=median($r_{ij}x_{ij}$) respectively. We set all $C_R$ and $C_A$ to unity since we wish to find effects that affect many (if not all) stars.

\item \textbf{convergence criteria:} The convergence criteria for the above iterations was unspecified in Tamuz \etal (2005). We define it as the iteration when the maximal absolute value of total correction $abs(A_j x_{ij} C_{A,i} + R_j C_{R,i})$ is smaller then some fraction $f$ of the standard deviation of that particular object. We used $f=0.5$ in our processing.

\item Once either additive or relative effects show no further correlation, one can use the regular Sysrem to look for additional effects of the other type since one may have different number of relative and additive effects --- until no effects of either type are identified. 

\item Bright stars both make planetary transits easier to detect, and are more susceptible to relative effects (that are later corrected by Sysrem/TFA (Kov{\'a}cs \etal 2005) /other). For these reasons some of the transit surveys intentionally monitor only the relatively bright stars in their field of view. On the other hand, fainter stars more readily show additive effects. We therefore propose to add more faint stars to the data of such surveys when using SARS as they might hold the key to better correct all stars.
\end{itemize}

\section{The SARS Post-Processing Scheme}
\label{SARSpipeline}

\subsection{Post-Processing Steps}

The above SARS core is just one element of the SARS post-processing scheme. We were able to achieve complete automation with no human input from \corot N2 FITS files to cleaned LCs. The post-processing global structure similar to the one used by MG09:
\begin{itemize}
\item Resample to 512s: resampling is done for each \corot color separately, if available.
\item Divide to $\sim$10d blocks, process blocks individually.
\item Subtract a running median with a window the size of 3 \corot orbital periods, and reject outliers
\item Choose a ``learning set" to calculate the effects with.
\item Apply the effects to all stars (we used three pairs of effects).
\item Re-set errors and reject bright outliers.
\end{itemize}

However, in order to achieve full automation we elaborate below on steps that were either unspecified or human-dependant on MG09:

\begin{itemize}
\item Outliers rejection is done in three tiers:
	\begin{enumerate}
	\item Removal of solitary outliers that are far from a small-window (5-point) median filter.
	\item Further outliers must meet to two criteria: 1) That frame has anomalously-high median-absolute-deviation (MAD - usually SAA-affected frames), 2) The data point is far from a 3-orbit median.
	\item Before SARS-core application, frames must have a minimum number of valid learning-set stars (we used at least 100).
	\end{enumerate}
\item Automatic choosing of the learning set aims to isolate the intrinsically- and instrumentally- constant stars. These stars are assumed to be numerous and similarly-variable in the raw data. An initial learning set is chosen by multiple criteria of:
	\begin{enumerate}
	\item The Alarm statistic of the LC (Tamuz \etal 2006) must be part of the bulk of Alarms. 
	\item The Alarm statistic of the residuals must be part of the bulk of Alarms.
	\item The locus of constant stars on the log(RMS) vs. Magnitude plot is along a straight line. Learning-set stars must not be far from that locus. 
	\end{enumerate}
	Next, the learning set is refined by a procedure inspired by techniques originally developed for photometric follow up of transiting planets (Holman \etal 2006) and is aimed at delivering the best comparison signal (lowest relative noise):
	\begin{enumerate}
	\item Given a set of N stars, calculate relative error on the total flux for all $N$ subsets of ($N-1$) stars.
	\item Compare the best subset (having the lowest relative flux error) with the relative flux error of the sum of $N$ stars:
		\item[-] If error reduced: repeat from step (i) with ($N-1$) stars
		\item[-] If error increased: optimal set reached
	\end{enumerate}
	This procedure guarantees that a local minimum in relative error is reached. We opt not to search for the global optimum since this is deemed too difficult (testing all subgroups of $N$ stars require testing $N!$ configurations -- were $N$ is in the order several thousands). We note that the resultant learning set preferentally includes faint stars (see Figure \ref{FractionVsMag}) which at least partially is because any variability is easier to spot on brighter stars. Interestingly, Figure \ref{FractionVsMag} and panel 2 of Figure \ref{SARSvsCLEANSET} show that despite the fact that faint stars are preferentially selected as learning-set stars -- bright stars are better corrected, showing that indeed something was learned from the fainter objects and was well applied to the brighter stars.

\begin{figure}
\includegraphics[width=0.5\textwidth]{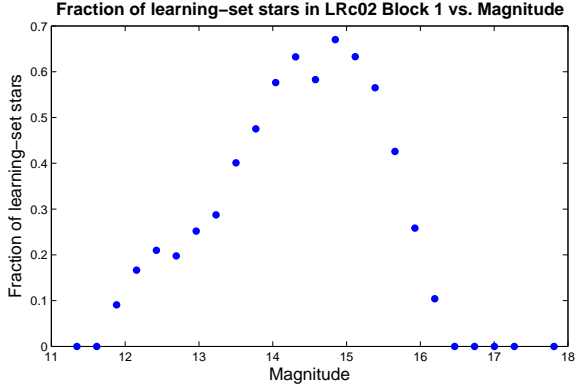}
\caption{A representetative plot of the fraction of learning-set stars in several magnitude bins (data here is from the example presented in \S \ref{Results}, for the learning set of the first effects-pair in the first block of LRc02).}
\label{FractionVsMag}
\end{figure}

\item We use the SARS core 3 times:
	\begin{enumerate}
	\item Use on the learning set only -- used to re-calibrate the errors only (see below).
	\item Use on the learning set only (now with calibrates errors) -- to calculate the effects.
	\item Use on the whole data set -- apply the already-calculated effect to all the LCs.
	\end{enumerate}
\item Errors re-setting is done by:
	\begin{equation}
	err_{ij}=StarErr(i) \frac{ExpErr(j)}{median(ExpErr)}
	\end{equation}
	Where StarErr is estimated from the star's LC and ExpErr is estimated from the distribution of magnitude residuals of each exposure.
\end{itemize}

\subsection{Results}
\label{Results}
\begin{figure}
\includegraphics[width=0.5\textwidth]{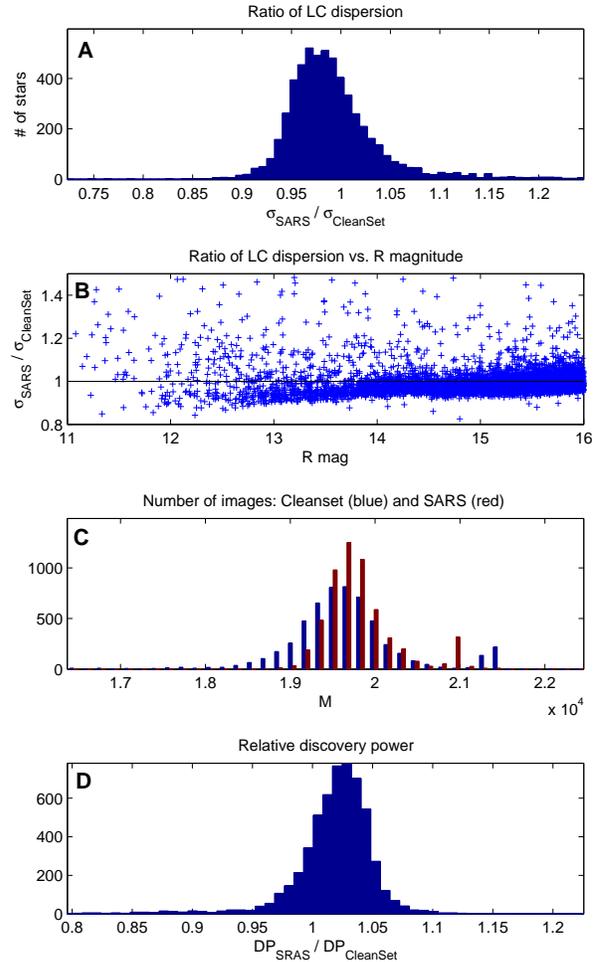}
\caption{Comparison between SARS and CleanSet for one random field (CCD E1 of LRc02). Panel A: histogram of the ratio of LC dispersion in SARS and CleanSet. Note $\sigma_{SARS} / \sigma_{CleanSet} < 1$ for most stars. Panel B: the above ratio of LC dispersion vs. R magnitude: the observed trend is explain in the text. Panel C: histogram of the number of remaining data points after outlier rejection in SARS and CleanSet. Panel D: histogram of the ratio of Detection Power (DP) between SARS and CleanSet}
\label{SARSvsCLEANSET}
\end{figure}

A comparison of the performance of SARS-cleaned and MG09-cleand LCs (the later sometimes dubbed "CleanSet") of one random field (LRc02) is shown in Figure \ref{SARSvsCLEANSET}: the SARS post processing delivers lower LC dispersion than in MG09's CleanSet for about $65\%$ of the stars, while keeping at least the same number of valid data points $M$ (CCD E2) if not more (CCD E1). If we compare the Detection Power, which is defined as:  $DP \sim \sqrt{M}/\sigma$, we find that it is higher in SARS than in CleanSet for up to $80\%$ of the stars. We note that there is a small trend in the relative performance: the brighter the stars are - the better SARS is relative to CleanSet. This is the expected result of the approximated functional form of the MG09 correction: since there are many more faint stars in the data than bright stars, the parabolic least-squares correction of MG09 tends to better suite the numerous faint ones, and so less fitting (due to the approximated functional form) to the bright stars. Thus our initial hypothesis that the effects are additive seems even more robust.

\begin{figure}
\includegraphics[width=0.5\textwidth]{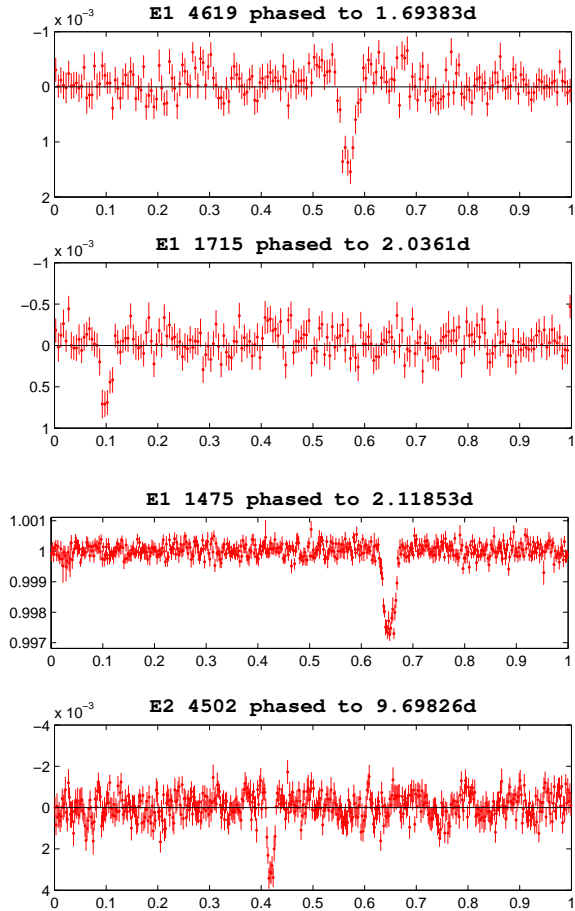}
\caption{Example of four \corot LRa02 candidates that were not detected by none of the detection teams before the application of the SARS post processing, and which are currently being followed up. The light curves are folded and binned to aid visibility.}
\label{LRa02}
\end{figure}

This global statistics is also translated to specific detections: so far we have SARS-analyzed three long runs (LRc02, LRa02, LRa01) and found about ten new transit-like signals that were not detected before in each field. For e.g., in Figure \ref{LRa02} we show four such LRa02 new transit candidates that were also chosen for follow-up. 

\section{Discussion}
\label{Discussion}

At the start of the \corot mission the \corot Exoplanets Science Team (CEST) made the strategic choice of having multiple team analyzing the exact same input data. By cross-checking each detection with different tools and cleaning techniques (e.g., Cabrera \etal 2009, Carpano \etal 2009) the CEST hoped to achieve the best possible transit candidates list for follow-up observations by the limited ground-based telescope resources. Here we present yet another step in the journey to clean photometric datasets in general and \corot data in particular: we generalize the popular and efficient Sysrem algorithm to include external parameters in a simultaneous solution, together with the unknown effects. This allows us to show that data from \corot is probably contaminated with additive, rather than relative, systematic effects -- and that these effects are the probable cause behind the phenomenological observation of MG09. The size of the additive correction $abs(A_j x_{ij} C_{A,i})$ is comperable to that of the relative correction $abs(R_j C_{R,i})$, with a median additive-to-relative ratio of about 0.5, but with a large scatter - making the additive correction larger than the relative correction for $\sim 1/3$ of the data points. Additive effects can arise from scattered light or erroneous bias or background subtraction, and we believe that the additive effects can be used just as the regular Sysrem effects to help to trace down the origin of the systematics and thus to avoid them in the first place.

We believe that the main advantage of SARS is not in a dramatic change in the standard deviation $\sigma$ of the LCs, but rather in the whiter color of their noise, which in turn allows for lower background of spurious signals in the BLS spectra (Kov{\'a}cs, Zucker \& Mazeh 2002) and thus the detection of shallower signals.

We were able to achieve good performance and complete automation which allows us to now process the entire mission data, and to look for -- and find -- ever shallower transit-like signals. For e.g., on LRc02 target \corot ID 0105842933 we were able to clearly detect a very shallow signal, only $~10^{-4}$ magnitudes deep, in a $P=1.08085 d$ period. Not only that, but we were also able to show that this is an eclipsing binary since at the double period the odd and even eclispes have different depths, with the secondary eclipse still visible (on a binned LC) while having a depth below the $84 ppm$ depth of an exo-Earth around a Sun-like star.

We will make the SARS-cleaned light curves available for the \corot community, and it is our intention that when the proprietary period is over to make data generally available (upon request). We note that we have also allowed for the application of SARS to the residuals of astrophysically variable stars (pulsators, eclipsing binaries, etc.) which will allow to better clean them too -- as part of a parallel CEST effort to look for transiting circumbinary planets (Ofir 2008, Ofir \etal 2009).

\label{lastpage}

\end{document}